\documentclass[]{dataflow}

\usepackage[toc,page,header]{appendix}

\usepackage{multirow}     
\usepackage{multicol}     
\usepackage{booktabs}     
\usepackage{colortbl}     
\usepackage{makecell}     
\usepackage{graphicx}     
\usepackage{float}        

\title{DataFlow-Harness: A Grounded Code-Agent Platform for Constructing Editable LLM Data Pipelines}

\author[*]{Runming He}
\author[*]{Zhen Hao Wong}
\author[*,\dagger]{Hao Liang}
\author[*]{Zimo Meng}
\author[]{Chengyu Shen}
\author[]{Xiaochen Ma}
\author[\ddagger]{Wentao Zhang}

\affiliation[]{$^{1}$Peking University, $^{2}$Institute for Advanced Algorithms Research, Shanghai, $^{3}$Zhongguancun Academy}

\abstract{
Large language models (LLMs) are increasingly used to automate data-processing workflows, yet coding agents typically produce scripts that are not automatically materialized as persistent, editable platform artifacts. We call this disconnect the \textit{NL2Pipeline gap}. To bridge it, we introduce \textsc{DataFlow-Harness}, a platform that guides an LLM agent to construct platform-native directed acyclic graphs (DAGs) through typed, incremental mutations rather than free-form scripts. The platform combines \textsc{DataFlow-Skills} for procedural guidance, a Model Context Protocol (MCP) layer that exposes the live operator registry and current pipeline state, and \textsc{DataFlow-WebUI}, which synchronizes conversational authoring with a visual DAG editor. On a 12-task data-engineering benchmark, \textsc{DataFlow-Harness} achieves a 93.3\% observed end-to-end pass rate. Relative to Vanilla Claude Code, it reduces measured monetary cost by 72.5\% and generation latency by 49.9\%; its observed pass rate is within 0.9 percentage points of the Context-Aware Claude Code baseline while its cost is 42.8\% lower. Per-task analysis indicates that Skills are most useful when construction depends on implicit procedural knowledge. These results show that live platform grounding can produce persistent, editable workflow artifacts with an observed reliability close to script-generation baselines and with lower measured construction cost and latency.

}

\date{\today}

\def\emailicon{\raisebox{-1.5pt}{\includegraphics[height=1.05em]{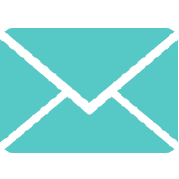}}}
\def\githubicon{\raisebox{-1.5pt}{\includegraphics[height=1.05em]{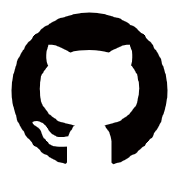}}}

\contribution[*]{Equal Contribution}
\contribution[\dagger]{Project Leader}
\contribution[\ddagger]{Corresponding author}

\checkdata[ \emailicon \hspace{0.3em} Correspondence ]{\email{wentao.zhang@pku.edu.cn}}

\newcommand{\sourcelink}{https://github.com/OpenDCAI/DataFlow-WebUI}

\checkdata[ \githubicon \hspace{0.3em} Source Code ]{ \url{\sourcelink} }

\checkdata[ Codebase Documentation ]{ \url{https://opendcai.github.io/DataFlow-Doc/} }

\begin{document}
\maketitle

\renewcommand{\thefootnote}{\fnsymbol{footnote}} 
\setcounter{footnote}{0}

\renewcommand{\thefootnote}{\arabic{footnote}}
\pagestyle{fancy}
\fancyhf{}
\fancyhead[L]{OpenDCAI Technical Report}
\fancyhead[R]{\thepage}

\newpage
\tableofcontents
\newpage

\section{Introduction}

Large language models (LLMs) are increasingly deployed to construct data-processing workflows for applications such as synthetic data generation, evaluation, retrieval augmentation, and model training~\cite{hong2025data, khattab2024dspy, gao2023retrieval}. Recent coding agents can automatically translate natural-language requirements into executable implementations, significantly reducing the effort required to construct such workflows~\cite{galster2026configuring,wu2024autogen}.

However, high task accuracy alone is often insufficient for production deployment. In industrial environments, workflow artifacts must remain visible, editable, reusable, and compatible with platform governance mechanisms throughout their lifecycle~\cite{kreuzberger2023machine}. In our experience, direct code-generation agents frequently produce disposable scripts that exist only as source code. These outputs are difficult to audit through graphical workflow interfaces and often hallucinate dependencies~\cite{patil2024gorilla, qin2024toolllm}, relying on unavailable operators, outdated platform assumptions, or framework-specific behaviors that general-purpose agents struggle to infer~\cite{liu2024agentbench, qin2024tool, qiao2023taskweaver}.

We define this challenge as the \textit{NL2Pipeline gap}: while users express workflow requirements in natural language, production environments require structured and persistent pipeline assets that can be visualized, edited, and reused. Here, a \emph{workflow} denotes the intended data-processing procedure, the \emph{pipeline representation} is its persistent platform object, and the \emph{DAG} captures its execution dependencies. Closing this gap requires more than improving code-generation accuracy: construction must remain grounded in platform semantics and produce artifacts that integrate with the host platform.

To address this limitation, we propose \textsc{DataFlow-Harness}, a platform for grounded workflow synthesis. Rather than directly generating scripts, \textsc{DataFlow-Harness} guides an agent to construct platform-native workflows through three decoupled components. \textsc{DataFlow-Skills} encode domain-specific construction knowledge, including operator-selection patterns, schema dependencies, and assembly procedures. The Model Context Protocol (MCP) provides access to the live operator registry and current workflow state, grounding agent actions in the execution environment~\cite{anthropic2024mcp}. Finally, \textsc{DataFlow-WebUI} provides a conversational interface for iterative refinement and materializes generated workflows as persistent, editable visual DAGs.

We evaluate \textsc{DataFlow-Harness} on pipeline-construction tasks covering data transformation, question answering, quality filtering, and synthetic data generation. Its observed end-to-end pass rate is close to the script-generation baselines on this benchmark, while measured token usage, cost, and latency are lower.

Our contributions are as follows:

\begin{itemize}
\item We formulate the \textit{NL2Pipeline gap}: the disconnect between natural-language workflow intent and persistent, platform-native workflow artifacts that remain available for inspection and editing.

\item We present \textsc{DataFlow-Harness}, which combines procedural Skills, live MCP grounding, typed incremental mutations, validation, and a synchronized conversational and visual authoring interface.

\item We evaluate reliability and construction efficiency on a 12-task benchmark, analyze where Skills help through a per-task ablation, and provide two controlled case studies of downstream training utility.

\end{itemize}

\section{Related Work}

\subsection{Agents for Code Generation.}
Code synthesis has evolved from foundational pretrained models such as Codex \citep{chen2021codex} and StarCoder \citep{li2023starcoder} to autonomous agentic loops. Reflexion \citep{shinn2023reflexion} and Self-Debug \citep{chen2023selfdebug} introduced iterative refinement via environment feedback. MCP \citep{anthropic2024mcp} provides unified tool interfaces for agents. Current agents such as SWE-agent \citep{yang2024swe} and Claude Code \citep{anthropic2024claude} focus on repository-level multi-turn editing and verification. Our work differs in that we constrain a general-purpose agent within a domain-specific harness rather than improving the agent itself.

\subsection{Data Engineering and LLM Pipelines.}
Data-centric AI \citep{zha2023datacentric} emphasizes systematic governance over model architecture. Specialized systems such as Data-Juicer \citep{chen2023datajuicer} and DCLM \citep{apple2024dclm} operationalize large-scale curation through extensible operators. DataFlow \citep{dataflow2025} and DSPy \citep{khattab2024dspy} treat LLM operations as composable components within formal execution graphs. Our work builds on DataFlow but focuses on agent-assisted construction of the pipeline itself, rather than pipeline execution.

\subsection{LLM-based Workflow Synthesis.}
Recent systems generate structured workflows rather than standalone programs. AutoFlow automatically synthesizes reusable workflows for LLM agents~\citep{li2024autoflow}, while Balis et al. translate scientific research questions into validated workflow DAGs and encode domain knowledge as reusable Skills~\citep{balis2026scientific}. These efforts establish the value of structured workflow generation. \textsc{DataFlow-Harness} focuses on a complementary systems problem: interactive, stateful authoring inside a live data-engineering platform. The agent retrieves the current pipeline and operator registry through MCP, applies typed incremental mutations to an existing artifact, validates each proposed state, and shares one persistent representation with the conversational interface and visual editor. Thus, our contribution is not NL-to-DAG generation alone, but platform-grounded construction and iterative editing of a live workflow artifact.

\section{System Architecture}
\label{sec:arch}

\begin{figure*}[t]
    \centering
    \includegraphics[width=0.9\textwidth]{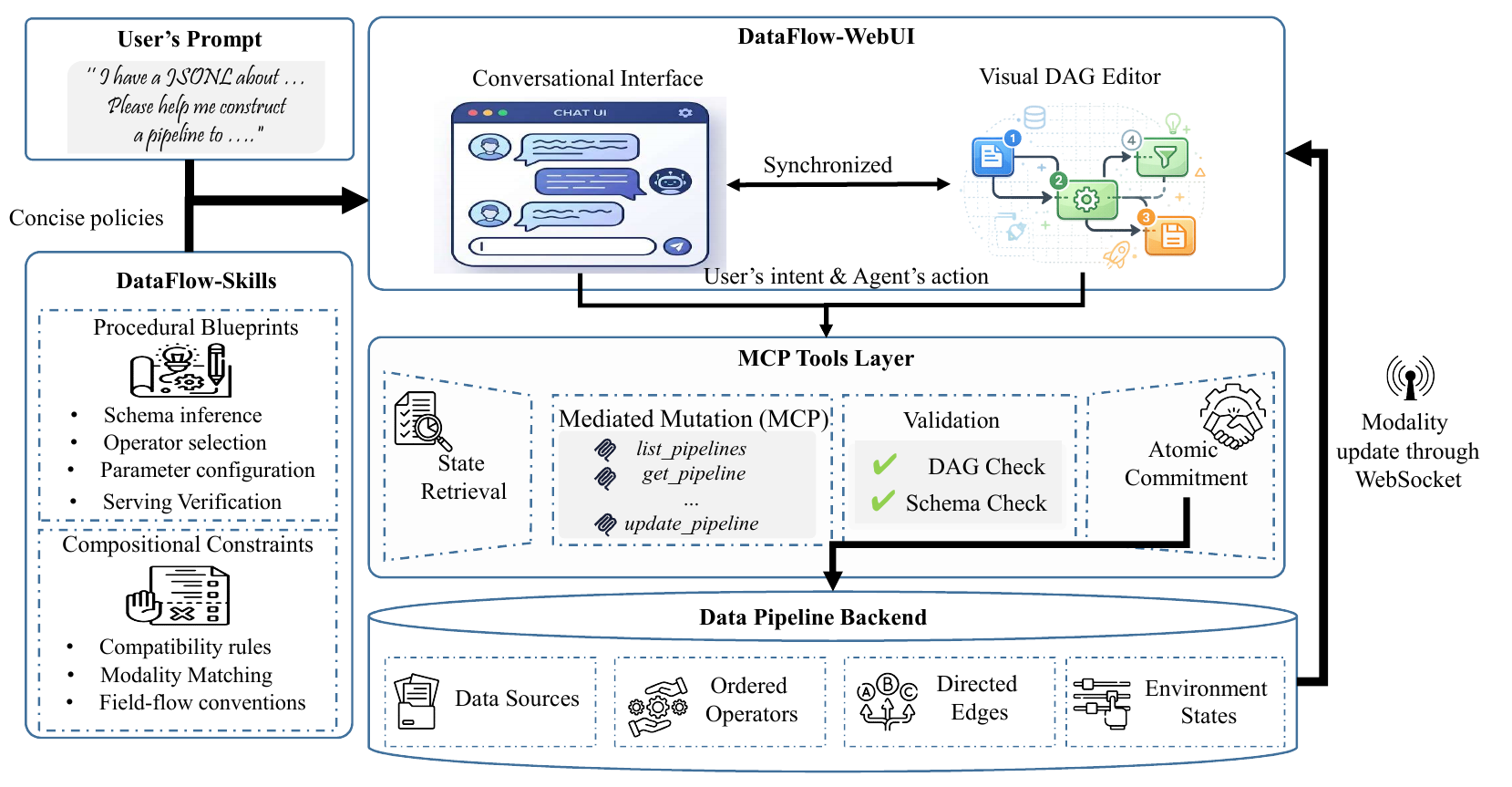}
    \caption{The \textsc{DataFlow-Harness} architecture. A shared pipeline representation is synchronized across the agent runtime and \textsc{DataFlow-WebUI}. \textsc{DataFlow-Skills} guides construction, while the Validation Engine checks DAG structure and schema compatibility.}
    \label{fig:system_archi}
\end{figure*}

\textsc{DataFlow-Harness} organizes workflow synthesis around four components: \textsc{DataFlow-WebUI}, the MCP Tools Layer, the Data Pipeline Backend, and \textsc{DataFlow-Skills}.

The operational lifecycle centers on the Data Pipeline Backend, which serves as the authoritative source of truth across conversational, visual, and programmatic interfaces. Pipeline mutations are issued through the MCP Tools Layer, validated and committed to the backend, and then synchronized with \textsc{DataFlow-WebUI}. In parallel, \textsc{DataFlow-Skills} provides procedural guidance that shapes agent reasoning without directly modifying pipeline state.

\subsection{Data Pipeline Backend}
\label{sec:formalism}
The Data Pipeline Backend serves as the authoritative source of truth for workflow synthesis. We represent a pipeline as $P=(D,O,E,S,R)$, where $D$ is the set of data sources and their URIs, $O$ is the set of configured operator instances, $E\subseteq O\times O$ contains directed data-dependency edges, $S$ records input and output field schemas, and $R$ contains runtime state such as model-serving endpoints.

Rather than generating free-form code, agents interact with the backend through typed mutations, including adding or removing operators, updating parameters, and connecting edges. A mutation is committed only if the resulting graph is acyclic and adjacent operator schemas are compatible. These checks establish structural validity; they do not by themselves guarantee semantic correctness, endpoint availability, or output quality.

\subsection{\textsc{DataFlow-WebUI}}

\begin{figure*}[t]
    \centering
    \includegraphics[width=0.98\textwidth]{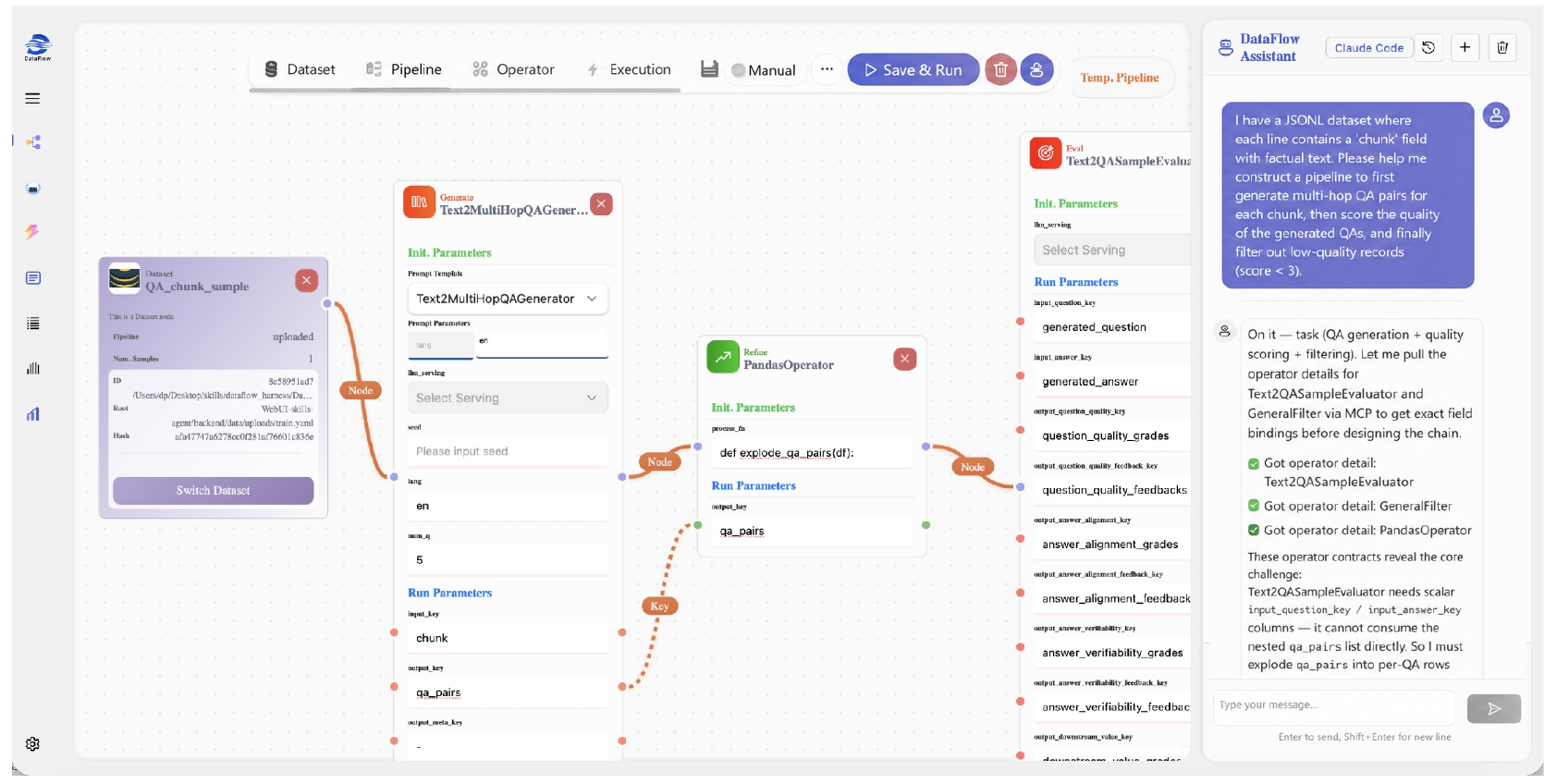}
    \caption{The dual-modality interface of \textsc{DataFlow-WebUI}, illustrating the synchronization between the conversational agent and the visual DAG editor.}
    \label{fig:interaction_ui}
\end{figure*}

\textsc{DataFlow-WebUI} provides two synchronized modalities for workflow construction: a conversational interface for natural-language authoring and a visual DAG editor for direct workflow inspection and editing (Figure~\ref{fig:interaction_ui}).

\paragraph{\textbf{Conversational Interface.}} Users describe workflow requirements in natural language. Before each agent turn, the current pipeline state and the DataFlow operator registry are injected into Claude Code's context via MCP. Guided by \textsc{DataFlow-Skills}, Claude Code interprets user intent and determines the required workflow modifications, which are expressed as MCP tool calls and applied to the Data Pipeline Backend.

\paragraph{\textbf{Visual DAG Editor.}} A graphical editor renders the workflow as a directed acyclic graph. Users can inspect agent-proposed changes, adjust parameters, relink edges, or add and remove operators directly. Any manual edit is immediately committed to the Data Pipeline Backend, ensuring that subsequent agent interactions operate on the latest workflow state without requiring explicit re-synchronization.

\subsection{MCP Tools Layer}

Every pipeline change, whether proposed by the agent or made manually, passes through a Request-Validate-Commit protocol.

\paragraph{\textbf{State Retrieval.}} At the start of each synthesis turn, the agent fetches the latest pipeline state, incorporating any manual edits since the previous turn.

\paragraph{\textbf{Mediated Mutation.}} Guided by \textsc{DataFlow-Skills}, Claude Code issues an MCP tool call expressing the intended change as a typed, structured mutation grounded in the DataFlow registry's live metadata.

\paragraph{\textbf{Validation.}} The system verifies two properties: that the updated pipeline remains a directed acyclic graph, and that the output field schema of each operator is compatible with the input schema of every downstream operator. Changes that fail either check are rejected.

\paragraph{\textbf{Validated Commitment.}} Validated changes are written to the backend store. A WebSocket notification broadcasts the updated state to connected clients, keeping the authoring modalities synchronized.

\subsection{\textsc{DataFlow-Skills}}
\label{sec:skills}

\textsc{DataFlow-Skills} provides procedural guidance for workflow synthesis by injecting domain-specific knowledge into Claude Code's reasoning context. While the MCP Tools Layer exposes operator metadata and workflow state, it does not encode recommended construction strategies or operator-composition best practices. As a result, agents may select inappropriate operators, omit prerequisite processing steps, or construct workflows that are structurally valid but semantically incorrect.

To address this, \textsc{DataFlow-Skills} encodes two classes of knowledge. The first consists of procedural blueprints that define recommended workflow-construction sequences, including schema inference, operator selection, parameter configuration, and serving verification. The second consists of compositional constraints that capture operator compatibility rules, such as modality matching and field-flow conventions for nested structures. Together, \textsc{DataFlow-Skills} guides agent reasoning, while the MCP Tools Layer grounds execution against the live DataFlow environment.

\section{Experiments}
\label{sec:eval}

We evaluate \textsc{DataFlow-Harness} to answer the following research questions, structured to progressively demonstrate the system's effectiveness, efficiency, and underlying mechanisms:

\textbf{RQ1 (Workflow Synthesis Effectiveness).} 
How does \textsc{DataFlow-Harness}'s native DAG synthesis compare to traditional free-form code generation (disposable scripts) in maintaining execution reliability and end-to-end task success for industrial data-processing tasks?

\textbf{RQ2 (System Efficiency).} 
What are the specific computational and economic advantages (e.g., token consumption, latency, and API cost) introduced by shifting from context-heavy script generation to our structured DAG synthesis?

\textbf{RQ3 (Ablation \& Micro-mechanisms).}
How does \textsc{DataFlow-Skills} improve upon pure tool-specification grounding (MCP-only) across varying levels of task complexity?

\textbf{RQ4 (Downstream Data Quality).}
Beyond governability and execution reliability, does grounding the agent with \textsc{DataFlow-Harness} lead it to author \emph{higher-quality} synthesis pipelines, as measured by the downstream accuracy of models trained on the data those pipelines produce?

\subsection{Experimental Setup}

\paragraph{\textbf{Benchmark.}}
We evaluate pipeline-construction capability on a benchmark of 12 tasks spanning six representative industrial data-processing scenarios: QA generation, review governance, long-document processing, multi-field scoring, schema normalization, and low-quality filtering. Each task specifies a natural-language objective together with input data samples and task-specific acceptance criteria. 

\paragraph{\textbf{Experimental Settings.}}

 To characterize the contributions of different system components, we compare four agent configurations with different levels of platform grounding: 
(1) \textbf{Vanilla CC}: An unconstrained coding baseline utilizing standard Claude Code. It relies entirely on internal parametric knowledge to generate disposable Python scripts, lacking access to platform-specific context. 
(2) \textbf{Context-Aware CC}: A repository-grounded baseline where the agent is provided with the raw DataFlow codebase. While in-context code comprehension enables it to accurately mimic platform operators, it inherently produces unmanageable, one-off scripts. 
(3) \textbf{MCP-Only}: A tool-augmented baseline strictly constrained to synthesize platform-native DAGs. It utilizes DataFlow MCP tools to dynamically discover operators but lacks explicit procedural guidance for complex assembly. 
(4) \textbf{\textsc{DataFlow-Harness}}: Our complete framework, combining MCP-based platform grounding with \textsc{DataFlow-Skills} to efficiently synthesize editable, governable workflow DAGs.

All experiments use Claude Opus 4.7 as the underlying large language model to keep the reasoning model fixed. To account for stochasticity in agent behavior and LLM generation, each task is executed 10 times under every setting, resulting in 120 task runs per method.

The configurations intentionally expose different action spaces: the script baselines may generate arbitrary Python, whereas MCP-only and \textsc{DataFlow-Harness} select and compose operators available in DataFlow. The comparison therefore measures the end-to-end system trade-off between free-form script generation and platform-constrained workflow construction, rather than an isolated difference in model capability.

\paragraph{\textbf{Evaluation Metrics.}}
We evaluate the proposed framework along three complementary dimensions: task success, efficiency, and platform integration.

\paragraph{\textbf{Task Success.}}
We measure \textbf{End-to-End (E2E) Pass}, which requires the generated workflow to execute successfully and produce outputs satisfying task-specific acceptance criteria. This metric captures overall workflow quality, including workflow synthesis, operator configuration, execution correctness, and final output validity.

\paragraph{\textbf{Efficiency Metrics.}}
To evaluate practical deployment cost, we additionally measure token consumption, monetary cost, and workflow construction latency. (1) \textbf{Token Consumption} reports the total number of input and output tokens used during workflow generation. (2) \textbf{Cost} is estimated using the official pricing of the underlying model and includes all interactions required to complete a workflow. (3) \textbf{Generation Latency} measures the wall-clock time from task submission to the production of a valid workflow artifact.

\subsection{Workflow Synthesis Effectiveness (\textbf{RQ1})}
\begin{table*}[t]
\centering
\small
\setlength{\tabcolsep}{6pt}
\begin{tabular}{ll|c|ccc}
\toprule
\multirow{2}{*}{\textbf{Method}} &
\multirow{2}{*}{\textbf{Artifact Type}} &
\textbf{Task Success (\%)} &
\multicolumn{3}{c}{\textbf{Efficiency \& Cost}} \\
\cmidrule(lr){3-3}
\cmidrule(lr){4-6}
&
& End-to-End Pass$\uparrow$
& Tokens (In/Out) $\downarrow$
& Cost (\$) $\downarrow$
& Latency (s) $\downarrow$ \\
\midrule

Vanilla CC
& Disposable Script
& 91.7
& 153,584 / 2,474
& 0.950
& 190.7 \\

Context-Aware CC
& Disposable Script
& \textbf{94.2}
& 185,626 / \underline{1,140}
& 0.456
& 115.9 \\

MCP-only
& Native DAG
& 83.3
& \underline{100,607} / 1,273
& \underline{0.321}
& \underline{105.5} \\

\textsc{DataFlow-Harness}
& Native DAG
& \underline{93.3}
& \textbf{74,958} / \textbf{891}
& \textbf{0.261}
& \textbf{95.5} \\

\bottomrule
\end{tabular}
\caption{
We report the end-to-end pass rate together with token usage, monetary cost, and generation latency. The pass rate is computed over 120 task runs (12 tasks $\times$ 10 trials); efficiency metrics are averaged over the same runs. \textbf{The best results are highlighted in bold, and the second-best results are underlined.}
}
\label{tab:eval}
\end{table*}

We first characterize the performance of free-form code generation. As shown in Table~\ref{tab:eval}, enriching vanilla generation (Vanilla CC) with execution context (Context-Aware CC) improves end-to-end success from 91.7\% to 94.2\%, highlighting the importance of procedural context for complex data-processing tasks. However, both approaches produce monolithic scripts that remain detached from platform-native workflow abstractions.

Transitioning from script generation to structured DAG synthesis introduces a substantial challenge. Using only operator specifications (MCP-only) reduces end-to-end success to 83.3\%, indicating that workflow constraints alone impose a significant reasoning burden on the model. This result reveals a clear \textit{NL2Pipeline gap}: directly generating deployable workflows is substantially harder than generating executable scripts.

\textsc{DataFlow-Harness} closes much of this gap through explicit procedural guidance. It achieves 93.3\% end-to-end success, improving by 10.0 percentage points over MCP-only while remaining within 0.9 percentage points of Context-Aware CC. The observed pass rates are numerically close, although we do not claim statistical equivalence. These results suggest that structured workflow synthesis can produce platform-native DAGs with reliability approaching that of the script-generation baselines on this benchmark.

\subsection{Efficiency and System Cost (\textbf{RQ2})}
Table~\ref{tab:eval} shows that \textsc{DataFlow-Harness} delivers substantial efficiency gains over free-form code generation. Compared with Vanilla CC, it reduces monetary cost by 72.5\% (from \$0.950 to \$0.261) and generation latency by 49.9\% (from 190.7s to 95.5s), while also achieving a higher end-to-end success rate. These results indicate that structured workflow synthesis is considerably more resource-efficient than generating executable scripts.

Notably, the efficiency gains persist even against the stronger Context-Aware CC baseline. Despite achieving nearly identical end-to-end performance, \textsc{DataFlow-Harness} reduces cost by 42.8\% and latency by 17.6\%, demonstrating a substantially more favorable efficiency--performance tradeoff.

The improvement is primarily driven by lower token consumption. Moving from script generation to native DAG synthesis, MCP-only nearly halves input token usage relative to Context-Aware CC, while \textsc{DataFlow-Harness} further reduces total token consumption by 25.5\% compared with MCP-only. This suggests workflow representations are far more compact than executable code, with procedural guidance further streamlining their construction within the constrained operator space.

\subsection{Textbook-to-VQA Workflow Evaluation}

We further evaluate our system on a challenging \textbf{textbook-to-VQA extraction} task, which requires constructing question-answer pairs from heterogeneous educational documents, including long-form textbooks, interleaved solution manuals, and exam answer sheets. This setting is particularly difficult due to (i) long-range dependencies between questions and answers, (ii) visually grounded reasoning over figures and tables, and (iii) highly non-linear document layouts that break local textual continuity.

To comprehensively evaluate extraction quality, we follow FlipVQA-Miner's~\cite{wong2025flipvqa} settings, and report two complementary metrics: \textbf{Precision} measures the correctness of extracted QA pairs, while
\textbf{Coverage Rate} measures the proportion of extractable QA pairs successfully recovered from the source document.

\begin{table}[t]
\centering
\small
\setlength{\tabcolsep}{6pt}
\begin{tabular}{l|c|c}
\toprule
Method & Precision $\uparrow$ & Coverage Rate $\uparrow$ \\
\midrule
Vanilla CC & 0.621 & 0.533 \\
Context-Aware CC & 0.893 & 0.801 \\
MCP-only & 0.784 & 0.621 \\
\textsc{DataFlow-Harness} & \textbf{0.972} & \textbf{0.873} \\
\bottomrule
\end{tabular}
\caption{Textbook-to-VQA extraction performance. Coverage measures the proportion of extractable QA pairs successfully recovered from the document.}
\label{tab:case_study}
\end{table}

Table~\ref{tab:case_study} evaluates a challenging textbook-to-VQA extraction task that requires composing document parsing, layout analysis, multimodal content extraction, question-answer alignment, and dataset construction into a single workflow. Compared with conventional workflow synthesis benchmarks, success in this setting depends not only on reasoning ability but also on the effective utilization of specialized document-processing components.

The observed results favor \textsc{DataFlow-Harness} on both extraction correctness and completeness, with 97.2\% precision and 87.3\% coverage. The largest absolute improvement is observed in coverage, where \textsc{DataFlow-Harness} recovers more valid QA pairs than the baselines. This pattern suggests that the framework constructs more complete workflows rather than merely filtering outputs conservatively. Confirming the generality of this result requires repeated runs and a fully specified annotation protocol.

The improvement may partly reflect the rich operator ecosystem provided by DataFlow. Textbook-to-VQA extraction requires capabilities such as PDF parsing, layout recovery, OCR, figure extraction, multimodal understanding, and long-range question-answer matching. While these capabilities already exist as reusable platform operators, effectively discovering and composing them remains challenging for general-purpose coding agents. The performance gap between MCP-only and \textsc{DataFlow-Harness} indicates that operator exposure alone is insufficient; procedural knowledge is required to guide workflow construction and ensure that relevant operators are assembled into valid end-to-end pipelines.

More broadly, this case study highlights an important property of \textsc{DataFlow-Harness}: its advantage does not arise from stronger model reasoning, but from enabling systematic reuse of mature platform assets. As workflow complexity increases, successful execution increasingly depends on leveraging existing operator ecosystems rather than synthesizing functionality from scratch. The results therefore provide concrete evidence that procedural guidance and platform-native abstractions are critical for closing the \textit{NL2Pipeline gap} in realistic data-processing scenarios.

\subsection{Ablation and Micro-mechanisms (\textbf{RQ3})}
Having shown that \textsc{DataFlow-Harness} largely closes the performance gap between native DAG synthesis and free-form code generation, we next investigate when explicit procedural guidance is most beneficial. The per-task results in Table~\ref{tab:per_task} reveal three consistent patterns.

\begin{table}[t]
\centering
\small
\setlength{\tabcolsep}{8pt}
\begin{tabular}{lcc}
\toprule
\textbf{Task} & \textbf{MCP-only} & \textbf{\textsc{DataFlow-Harness}} \\
\midrule

\multicolumn{3}{l}{\emph{Procedural-knowledge-dependent tasks}} \\
1a: QA basic            & 6/10 & \textbf{10/10} \\
1b: QA with filter      & 6/10 & \textbf{9/10} \\
3b: Text-to-QA chain    & 6/10 & \textbf{10/10} \\
\midrule

\multicolumn{3}{l}{\emph{Trivially routable tasks}} \\
5a: Field rename         & 10/10 & 10/10 \\
5b: Nested flatten       & 10/10 & 10/10 \\
6a: Length filter        & 10/10 & 10/10 \\
6b: LLM semantic filter  & 10/10 & 10/10 \\
\midrule

\multicolumn{3}{l}{\emph{Tasks with non-synthesis bottlenecks}} \\
4a: Score and filter        & 7/10 & 7/10 \\
4b: Multi-dimensional score & 7/10 & 7/10 \\
2a: Sentiment               & 9/10 & \textbf{10/10} \\
2b: Review governance       & \textbf{10/10} & 9/10 \\
3a: Long-document summary   & 9/10 & \textbf{10/10} \\
\bottomrule
\end{tabular}

\caption{Per-task end-to-end pass counts over 10 independent trials. Tasks are grouped according to the mechanisms discussed in RQ3.}

\label{tab:per_task}
\end{table}

\paragraph{Procedural guidance is most valuable when task success depends on implicit domain knowledge.}
The largest improvements occur on QA-generation and language-processing tasks (\texttt{1a}, \texttt{1b}, and \texttt{3b}), where successful construction requires more than selecting compatible operators. Although MCP-only frequently generates structurally valid DAGs, it struggles to infer task-specific procedures from operator descriptions alone. Encoding these procedures as reusable skills improves aggregate success in this group from 18/30 to 29/30 runs, accounting for most of the overall gain.

\paragraph{Procedural guidance provides limited benefit when workflow routing is straightforward.}
For simple transformation and filtering tasks (\texttt{5a}, \texttt{5b}, \texttt{6a}, and \texttt{6b}), both methods achieve perfect success. In these cases, operator specifications are sufficient to identify the workflow structure, leaving little room for improvement from higher-level guidance.

\paragraph{Procedural guidance cannot overcome limitations outside workflow synthesis.}
The remaining failures appear to arise primarily from factors unrelated to workflow construction. Both methods exhibit the same failure rate on multi-field scoring tasks (\texttt{4a} and \texttt{4b}), where outputs occasionally violate downstream numerical constraints despite correct DAG generation. On tasks \texttt{2a}, \texttt{2b}, and \texttt{3a}, differences are small and occasionally favor MCP-only, suggesting that prescriptive procedures may reduce flexibility when multiple execution strategies are valid.

Overall, these results indicate that \textsc{DataFlow-Skills} contribute primarily by injecting procedural knowledge that is difficult to recover from operator specifications alone. Their observed benefit is largest on ambiguous, procedurally complex tasks and diminishes when workflow construction is trivial or bottlenecked by the underlying model. Because this ablation compares MCP-only with the full system, it does not separately identify the contribution of validation.

\subsection{Downstream Training Utility Evaluation (\textbf{RQ4})}
\label{sec:rq4}

The preceding experiments assess whether \textsc{DataFlow-Harness} produces workflows that are governable and that execute reliably. A distinct and arguably more consequential question is whether the harness helps the agent author \emph{better} pipelines, i.e., pipelines whose \emph{output data} is more useful downstream. To answer this, we adopt an end-to-end, outcome-based protocol: we let the coding agent construct a full data-synthesis pipeline, run the pipeline to synthesize a training set, fine-tune a model on that set, and measure the resulting model's benchmark accuracy. Because a pipeline is ultimately a means to produce training data, the quality of the data it emits is a direct, if indirect, measure of the pipeline's quality.

\paragraph{\textbf{Protocol.}}
For each scenario we compare two configurations that differ in exactly one respect. In \textbf{Vanilla CC}, Claude Code receives only the natural-language task description and writes a synthesis pipeline from its parametric knowledge. In \textbf{\textsc{DataFlow-Harness}}, the identical agent receives the identical prompt but is additionally grounded in the DataFlow environment through \textsc{DataFlow-Skills} and the MCP operator registry. Everything downstream of pipeline construction is held fixed across the two arms: the same underlying LLMs and OpenAI-style API settings (concurrency and timeout) are used to synthesize data, the same number of samples is generated, and the same base model, fine-tuning recipe, and evaluation harness are used. This controlled setup isolates the agent-authored pipeline as the primary source of differences in downstream accuracy, while holding model, data scale, and training settings fixed.

\paragraph{\textbf{Math Reasoning Pipeline.}}
The first scenario (Prompt~1) asks the agent to build a math data cleaning-and-synthesis pipeline: verify and filter incoming problems, discard ill-posed items, expand each seed question into two new synthetic problems, generate reasoning traces for every question, and apply $n$-gram deduplication to the resulting QA pairs, with per-stage models specified separately. We run each agent-authored pipeline over the same seed pool, fine-tune Qwen2.5-32B-Instruct following the LIMO recipe~\cite{ye2025limo} (full-parameter SFT, lr $5\mathrm{e}{-6}$, cosine schedule without warmup, batch size $64$, 16K context), and evaluate on the math suite and protocol of DataFlow~\cite{dataflow2025} (Table~5), reporting AIME24/25 as avg@32.

\begin{table*}[t]
\centering
\small
\setlength{\tabcolsep}{5pt}
\caption{\textbf{Math pipeline quality via downstream training.} Both pipelines are authored by Claude Code from the \emph{same} natural-language prompt (Prompt~1), synthesize data with the \emph{same} models and API settings, and are used to fine-tune Qwen2.5-32B-Instruct under the \emph{same} LIMO recipe. We report accuracy (\%) on the DataFlow~\cite{dataflow2025} Table~5 math suite; AIME24/25 are reported as avg@32. At matched epochs, data produced with \textsc{DataFlow-Harness} yields a higher average, indicating a higher-quality pipeline. Per-epoch best average is in \textbf{bold}.}
\label{tab:cc_math}
\resizebox{\textwidth}{!}{
\begin{tabular}{lccccccccc}
\toprule
\textbf{Pipeline (author)} & \textbf{GSM8K} & \textbf{MATH} & \textbf{AMC23} & \textbf{Olympiad} & \textbf{Gaokao24\_mix} & \textbf{Minerva} & \textbf{AIME24@32} & \textbf{AIME25@32} & \textbf{Avg} \\
\midrule
\textbf{Qwen2.5-32B-Instruct (base)} & 95.8 & 73.5 & 70.0 & 38.5 & 42.9 & 26.5 & 16.8 & 11.6 & 46.95 \\
\midrule
\rowcolor[rgb]{.867, .922, .969}
\multicolumn{10}{c}{\textit{\textbf{Trained with 1 epoch}}} \\
\midrule
\quad Vanilla CC                    & 92.3 & 78.0 & 47.5 & 42.8 & 56.0 & 35.7 & 25.1 & 21.6 & 49.9 \\
\quad \textsc{DataFlow-Harness}     & 93.9 & 72.3 & 72.5 & 38.7 & 38.5 & 26.5 & 35.9 & 34.5 & \textbf{51.6} \\
\midrule
\rowcolor[rgb]{.867, .922, .969}
\multicolumn{10}{c}{\textit{\textbf{Trained with 2 epochs}}} \\
\midrule
\quad Vanilla CC                    & 94.8 & 84.0 & 60.0 & 48.0 & 53.8 & 39.7 & 31.8 & 24.3 & 54.5 \\
\quad \textsc{DataFlow-Harness}     & 94.4 & 76.6 & 75.0 & 45.2 & 42.9 & 25.7 & 45.4 & 40.0 & \textbf{55.7} \\
\bottomrule
\end{tabular}
}
\end{table*}

Table~\ref{tab:cc_math} reports the results at matched training budgets. Both configurations lift the base model substantially, confirming that Claude Code can author a functional synthesis pipeline unaided. However, at every matched epoch the data produced under \textsc{DataFlow-Harness} yields a higher average accuracy, improving from $49.9$ to $51.6$ after one epoch and from $54.5$ to $55.7$ after two. The gains concentrate on the hardest, most contamination-sensitive benchmarks: \textsc{DataFlow-Harness} data raises AIME24@32 from $25.1$ to $35.9$ and AIME25@32 from $21.6$ to $34.5$ at one epoch, with a similar margin at two epochs. This pattern is consistent with the grounded pipeline applying more effective verification, filtering, and deduplication, thereby producing cleaner and more challenging reasoning traces rather than merely more of them.

\FloatBarrier

\paragraph{\textbf{General SFT Pipeline.}}
The second scenario (Prompt~2) is more demanding: the agent must build a \emph{from-scratch} general instruction-tuning pipeline with no seed dataset, generating data end-to-end through API calls. The pipeline spans three stages: topic-conditioned generation of diverse instruction--response pairs from a preset knowledge taxonomy with multiple difficulty levels; a critique-then-rewrite refinement pass; and an LLM-as-judge scoring stage that filters low-quality samples. Each agent-authored pipeline synthesizes 10K instruction--response pairs, which we use to fine-tune Qwen2.5-7B-Base under an identical recipe (full-parameter SFT with LLaMA-Factory and DeepSpeed ZeRO-3 on 8$\times$H20, lr $1\mathrm{e}{-5}$, cosine schedule, warmup $0.03$, 3 epochs, global batch $128$, cutoff $4096$, bf16, seed $42$). We evaluate the fine-tuned models with lm-evaluation-harness~\cite{eval-harness} across knowledge (MMLU), math (GSM8K, MATH, Minerva, Olympiad), and code (HumanEval, MBPP and their EvalPlus~\cite{liu2023evalplus} variants) benchmarks.

\begin{table}[H]
\centering
\small
\setlength{\tabcolsep}{5pt}
\caption{\textbf{General SFT pipeline quality via downstream training.} Both pipelines are authored by Claude Code from the \emph{same} from-scratch synthesis prompt (Prompt~2) and generate 10K instruction--response pairs with the \emph{same} models and API settings. We fine-tune Qwen2.5-7B-Base under an identical recipe (full-parameter SFT, LLaMA-Factory + DeepSpeed ZeRO-3, 8$\times$H20, lr $1\mathrm{e}{-5}$, cosine, warmup $0.03$, 3 epochs, global batch $128$, cutoff $4096$, bf16, seed $42$) and evaluate with lm-evaluation-harness~\cite{eval-harness}. Code benchmarks use the EvalPlus~\cite{liu2023evalplus} variants (HE+/MBPP+). \textbf{Avg} is the mean over all nine benchmarks; best per column in \textbf{bold}.}
\label{tab:cc_general}
\resizebox{\textwidth}{!}{
\begin{tabular}{lc|cccc|cccc|c}
\toprule
& \textbf{Knowledge} & \multicolumn{4}{c|}{\textbf{Math}} & \multicolumn{4}{c|}{\textbf{Code}} & \\
\cmidrule(lr){2-2} \cmidrule(lr){3-6} \cmidrule(lr){7-10}
\textbf{Pipeline (author)} & \textbf{MMLU} & \textbf{GSM8K} & \textbf{MATH} & \textbf{Minerva} & \textbf{Olympiad} & \textbf{HumanEval} & \textbf{HE+} & \textbf{MBPP} & \textbf{MBPP+} & \textbf{Avg} \\
\midrule
Vanilla CC                & \textbf{74.4} & \textbf{82.9} & 68.2 & 27.6 & 35.9 & 78.0 & 70.1 & 64.6 & 51.6 & 61.5 \\
\textsc{DataFlow-Harness} & 74.2 & 79.5 & \textbf{70.1} & 27.6 & \textbf{36.3} & \textbf{80.5} & \textbf{72.6} & \textbf{75.4} & \textbf{58.2} & \textbf{63.8} \\
\bottomrule
\end{tabular}
}
\end{table}

Table~\ref{tab:cc_general} shows nearly identical knowledge performance (MMLU $74.2$ vs.\ $74.4$), while the two pipelines trade wins across the individual math benchmarks. The clearest difference emerges on code, where the \textsc{DataFlow-Harness} pipeline is stronger across all four benchmarks, with the largest gap on MBPP ($75.4$ vs.\ $64.6$). These code gains lift the overall nine-benchmark average by $2.3$ points ($63.8$ vs.\ $61.5$). This pattern is consistent with the grounded critique-then-rewrite and judge stages producing more executable, better-structured coding responses. Together, the two scenarios provide preliminary outcome-level evidence that grounding can improve the utility of data produced by an agent-authored pipeline. Because each scenario is a controlled case study rather than a repeated experiment across independently authored pipelines and multiple training seeds, these results should not be interpreted as a general causal estimate.

\FloatBarrier

\section{Conclusion}
We presented \textsc{DataFlow-Harness}, a platform that addresses the \textit{NL2Pipeline gap} by combining procedural Skills, live MCP grounding, typed mutations, structural validation, and synchronized conversational and visual editing. On our 12-task benchmark, its observed end-to-end pass rate is close to the script-generation baselines, while measured construction cost and latency are lower. The per-task ablation further shows where procedural guidance is most useful.

\paragraph{Limitations.}
Our evaluation uses one coding-agent and model family and a relatively small, platform-specific benchmark. The available ablation does not isolate every component, and schema validation cannot guarantee semantic correctness. We report observed averages without task-clustered confidence intervals or a pre-specified non-inferiority test. Cost reporting also requires a token-class breakdown to be independently recomputed when prompt caching is used. Finally, the downstream-utility results cover two case studies without multiple independently authored pipelines and training seeds. Direct evaluation of persistence, reuse, provenance, concurrent editing, and recovery is needed before making broader workflow-governance claims.

\bibliographystyle{plainnat}
\bibliography{main}




\end{document}